\documentclass[a4paper,twocolumn,showpacs,amsfonts,aps,prl,floatfix,nofootinbib]{revtex4-1}
\usepackage{graphicx}
\usepackage{amsfonts}
\usepackage{amsmath}
\usepackage{amsbsy}
\usepackage{longtable}
\usepackage{bm}
\usepackage{hyperref}
\usepackage{float}

\begin{document}

\title{Dissipative hydrodynamics for relativistic multi-component systems}

\author{Andrej El$^1$}\email{el@th.physik.uni-frankfurt.de}
\author{Ioannis Bouras$^1$}
\author{Francesco Lauciello$^1$}
\author{Zhe Xu$^{1,2}$}
\author{Carsten Greiner$^1$}

\affiliation{$^1$Institut f\"ur Theoretische Physik, Goethe-Universit\"at, Max-von-Laue-Str.\ 1,
D-60438 Frankfurt am Main, Germany\\
$^2$Frankfurt Institute for Advanced Studies,
Ruth-Moufang-Strasse\ 1,
D-60438 Frankfurt am Main, Germany}

\begin{abstract}
Second-order dissipative hydrodynamic equations for each component of
a multi-component system are derived using the entropy principle.
The shear viscosity of the whole system, appearing in the equation summed-up 
 over all components, is related to the partial shear pressures and cannot be
considered as an external parameter. We demonstrate that it is essential to solve 
hydrodynamic equations for each component, instead of treating a mixture
as an effective one-component system with a free parameter $\eta/s$.
Thus, extractions of the $\eta/s$ value of the QGP at RHIC and LHC have to be
reexamined.
\end{abstract}


\date{\today}

\maketitle

The deconfined state of QCD matter produced at the early stage of 
ultrarelativistic heavy-ion collisions at RHIC and LHC is a multi-component
system with quark and gluon degrees of freedom.
Large values of elliptic flow coefficient $v_2$, observed in experiments 
of ultrarelativistic heavy-ion collisions
at RHIC \cite{Adler:2003kt,Voloshin:2008dg} and LHC \cite{Aamodt:2010pa},
indicate that the produced quark-gluon plasma (QGP) is a nearly perfect
fluid. This has motivated rapid developments on relativistic dissipative 
hydrodynamic formalisms 
\cite{IS,Muronga:2003ta,Romatschke:2009im,Denicol:2010xn,Monnai:2010qp}. The value of the shear
viscosity to the entropy density ratio $\eta/s$ for the QGP at RHIC and LHC was extracted from
comparisons of hydrodynamic \cite{hydros} as well as kinetic transport \cite{Xu:2007jv} calculations
with experimental data. All these hydrodynamic formalisms
are based on the assumption that the quark-gluon mixture can be regarded as an effective
one-component system, where $\eta/s$ is an external parameter characterizing
the dissipation in the system. One may ask the question whether this
assumption really holds in general, which is of major interest for investigation of the QGP
properties.

In this Letter we demonstrate that a standard one-component hydrodynamic description with a single
shear viscosity coefficient in general cannot be applied to a multi-component system. We will
explain this statement by deriving second-order
dissipative hydrodynamic equations for a multi-component system from the entropy principle. Our
approach differs from the one reported in Ref. \cite{Monnai:2010qp}, since we introduce separate
evolution equations and transport coefficients for each component of the mixture. We then
show that by summing-up equations for all components one can obtain an equation for the system as a
whole, which has a relaxation-type form characteristic for all the present hydrodynamic formalisms,
but the effective shear viscosity for the mixture is now related to the partial shear pressures of
its
components and thus cannot be considered as an externally specified parameter. It is essential to
solve the
hydrodynamic equations for each component. We will confirm our findings by comparing the solutions
of the hydrodynamic equations with those from kinetic transport calculations.

We consider a mixture of $N$ particle species. Neglecting bulk pressure and
heat flow we construct the total entropy current as \cite{Kranys}
\begin{equation}
\label{smu1}
s^\mu=\sum_{i=1}^N s_i^\mu=s_{eq} u^\mu - \sum_{i=1}^N \frac{\beta_i}{2T_i}
\pi_{i,\alpha\beta} \pi_i^{\alpha\beta} u^\mu \,,
\end{equation}
where $s_{eq}$ is the total entropy density in local equilibrium and $u^\mu$ 
is the hydrodynamic velocity. $T_i$ and $e_i$ are the temperature and
local energy density of the particle species $i$. $\beta_i=(9/4e_i)$
and $\pi_i^{\mu\nu}=T_i^{\mu\nu}-T_{i,eq}^{\mu\nu}$ is the shear stress tensor,
which is the difference between the energy-momentum tensor $T_i^{\mu\nu}$ 
and the equilibrium one. Equation (\ref{smu1}) is the generalization
of the entropy current for a one-component system ($N=1$).

The total entropy production is then
\begin{equation}
\partial_\mu s^\mu = \sum_i \pi_{i,\alpha\beta} 
\left[ \frac{\sigma^{\alpha\beta}}{T_i} - 
\pi_i^{\alpha\beta} \partial_\mu \left( \frac{\beta_i}{2T_i} u^\mu \right) -
\frac{\beta_i}{T_i} u^\mu \partial_\mu \pi_i^{\alpha\beta} \right] \,,
\label{dmusmu1}
\end{equation}
where
\begin{equation}
\sigma^{\mu\nu}=\nabla^{\langle\mu}u^{\nu\rangle}=\left(\frac{1}{2}(\Delta_{\alpha}^{\mu}\Delta_{\beta}^{\nu}+\Delta_{\alpha}^{\nu}\Delta_{\beta}^{\mu})-\frac{1}{3}\Delta_{\alpha\beta}\Delta^{\mu\nu} \right)\nabla^{\alpha}u^{\alpha}
\nonumber
\end{equation}
and $\Delta_{\alpha\beta}=g_{\alpha\beta}-u_\alpha u_\beta$ with
the metric $g_{\alpha\beta}={\rm diag} (1,-1,-1,-1)$. We have used conservation of the partial
particle flows and total energy-momentum tensor, $\partial_\mu N_i^\mu=0$ and $\partial_\mu
T^{\mu\nu}=0$, to obtain Eq. (\ref{dmusmu1}).

According to the second law of thermodynamics the entropy production
is non-negative. A simple way to fulfill this is to make the terms in the
square bracket in Eq. (\ref{dmusmu1}) to be proportional to 
$\pi_i^{\alpha\beta}$, $[\cdots ]=\pi_i^{\alpha\beta}/(2\eta_i T_i)$.
Then the entropy production (\ref{dmusmu1}) has the following algebraic 
structure:
\begin{equation}
\partial_\mu s^\mu \overset{!}{=} \sum_{i=1}^N
\frac{\pi_{i,\alpha\beta}\pi_i^{\alpha\beta}}{2\eta_i T_i} \ge 0 \,.
\label{dmusmu2}
\end{equation}
This leads to the dynamical equation for each $\pi_i^{\alpha\beta}$:
\begin{equation}
\label{pi1}
u^\mu \partial_\mu \pi_i^{\alpha\beta} = 
-\frac{\pi_i^{\alpha\beta}}{2\eta_i\beta_i}
-\pi_i^{\alpha\beta} \frac{T_i}{\beta_i}
\partial_\mu\left(\frac{\beta_i}{2T_i} u^\mu\right) 
+\frac{\sigma^{\alpha\beta}}{\beta_i} \,,
\end{equation}
which was introduced by Israel and Stewart for a one-component system
($N=1$) \cite{IS,Muronga:2003ta}.
The main subject of this Letter is to confirm the validity of
Eq. (\ref{pi1}) for a multi-component system ($N>1$).
Before we do so, we shall first determine the coefficients $\eta_i$, 
which in general differ from the usual definition of the shear viscosity.
The reason for this is that $\pi_i^{\alpha\beta}$'s are correlated due to
interactions between particles from different species.
These correlations between $\pi_i^{\alpha\beta}$'s can only be seen, when
each $\eta_i$ depends on all $\pi_j^{\alpha\beta}$, $j=1,2, \cdots, N$.
We will also show later that the $\eta_i$ become the shear viscosities, when
the ratios of components of $\pi_i^{\alpha\beta}$'s are relaxing to constants in time.

We now make use of relativistic kinetic theory and express the entropy current 
via the particle distribution function in phase space $f_i(x,p_i)$:
\begin{equation}
\label{smu2}
s^\mu=\sum_{i=1}^N \int d\Gamma_i \ p_i^\mu f_i(x,p_i)\, [1-\ln f_i(x,p_i)]
\end{equation}
with $d\Gamma_i=d^3p_i/E_i/(2\pi)^3$. It was shown for the case of 
$N=1$ \cite{3rdO} and is obviously true for $N>1$ that using the Grad's 
ansatz \cite{Muronga:2003ta}
$f_i(x,p_i) = f_{i,eq}(x,p_i) ( 1 + A_i \pi_{i,\mu\nu} p_i^\mu p_i^\nu)$
in Eq. (\ref{smu2}) one obtains Eq. (\ref{smu1}) up to second order
in $\pi_{i,\mu\nu}$. Here $f_{i,eq}(x,p_i)$ is the equilibrium distribution
function and $A_i= [ 2 (e_i + P_i) T_i^2 ]^{-1}$, where $P_i$ is the pressure.

Assuming that space-time evolutions of $f_i(x,p_i)$ obey the Boltzmann 
equations
\begin{equation}
\label{boltzmann}
p_i^\mu \partial_\mu f_i = C_i[f_1, f_2, \cdots, f_N]= 
C_{ii}[f_i] +  \sum_{j=1,j\ne i}^N C_{ij}[f_i,f_j] \,,
\end{equation}
where $C_{ii}$ are the collision terms describing interactions of
particles of same species and $C_{ij}$ describing binary interactions of
particles of different species. Explicit expressions for the collision 
terms can be found for example in \cite{Xu:2004mz}. Taking derivative 
of (\ref{smu2}) and using (\ref{boltzmann}) we obtain
\begin{equation}
\label{dmusmu3}
\partial_\mu s^\mu = \sum_{i=1}^N A_i \pi_{i,\mu\nu} 
\int d\Gamma_i \, p_i^\mu p_i^\nu C_i \,.
\end{equation}
Comparison between Eqs. (\ref{dmusmu3}) and (\ref{dmusmu2}) leads to
\begin{equation}
\eta_{i} = \frac{\pi_{i,\mu\nu} \pi_i^{\mu\nu}}
{2A_i \pi_{i,\mu\nu} \int d\Gamma_i p_i^\mu
p_i^\nu  C_i} \,.
\label{eta_gen}
\end{equation}
Because the collision term $C_i$ is a functional of all $f_j$'s,
each $\eta_i$ depends on all $\pi_j^{\mu\nu}$'s with $j=1,2,\cdots,N$.

Equations (\ref{pi1}) together with (\ref{eta_gen}) are the main findings
in this work. They provide the dynamic evolution of the shear pressure 
for each component of the mixture. This is highly relevant for the present
research on high energy nucleus-nucleus collisions at RHIC and LHC,
where the produced quark-gluon plasma is a mixture, and thermal equilibration of 
quarks and gluons is expected to proceed differently during the
expansion. For a one-component system, the shear viscosity $\eta$,
or practically the shear viscosity to the entropy density ratio $\eta/s$,
can be regarded as a to be specified but free external parameter. In contrast to that, for a
multi-component system we recognize that each transport coefficient $\eta_i$
depends on shear pressures of all components and, thus, is not a free 
parameter. This is a new property, which is not observed for 
a one-component system, although the basis
of the derivation of Eqs. (\ref{pi1}) is the same as in the second-order
Israel-Stewart theory for a one-component system, namely the entropy
principle Eq. (\ref{dmusmu2}). One may ask the question whether it is
still correct to regard a multi-component mixture as an effective
one-component system with a freely chosen $\eta/s$, as mostly done in 
Refs. \cite{hydros} for the quark-gluon plasma.

In the rest of this Letter we will discuss some applications of the obtained equations. To
confirm our findings, we solve the multi-component hydrodynamic equations and compare the results
with those from kinetic transport theory.

For reasons of simplification we consider a one-dimensional expansion with the 
boost-invariance. In this case $u^\mu=(t,0,0,z)/\tau$, where 
$\tau=\sqrt{t^2-z^2}$ is the proper time. Because of the transversal 
isotropy the shear stress tensor is then 
$\pi_i^{\mu\nu}=diag(0,\pi_i/2,\pi_i/2,-\pi_i)$ in the co-moving frame. $\pi_i$ denotes the shear
pressure of species $i$. The hydrodynamic equations (\ref{pi1}) are reduced to
\begin{equation}
{\dot \pi}_i = -\frac{2}{9}\frac{\pi_ie_i}{\eta_i}
-\frac{4}{3}\frac{\pi_i}{\tau} +
\frac{8}{27}\frac{e_i}{\tau}  \,.
\label{dotpi}
\end{equation}
$\dot{\pi}_i$ is the derivative with respect to $\tau$.
The set of equations is the multi-component analog of the known 
Israel-Stewart second-order 
equations \cite{IS,HM09,Muronga:2003ta,3rdO} 
and has a relaxation-type form, existence of which was advocated 
in Ref. \cite{Denicol:2011fa}.

Furthermore, we consider only binary elastic collisions, which keep 
the particle number of each species conserved. Particles are assumed
to be massless Boltzmann particles. We also take an isotropic
distribution for the collision angle. With these simplifications
we can analytically calculate the integrals in Eq. (\ref{eta_gen}) 
and obtain (for details see \cite{El:prep})
\begin{equation}
\eta_{i}^{-1}  = T_i^{-1} \sum_{j=1}^N \left (\frac{7}{6}\frac{n_j}{n_i}
- \frac{1}{3}\frac{\pi_j}{\pi_i} \right ) \sigma_{ij} \,,
\label{eta_iso} 
\end{equation}
where $n_i$'s are the local particle number densities, $\sigma_{ij}$
denotes the cross section for a collision between a particle of species $i$ and a particle of
species $j$, and $\sigma_{ii}$ is the cross section for a collision of two identical
particles of species $i$. One can clearly see that $\eta_i$ explicitly depends not only on
temperature, cross sections, and the chemical
composition, but also on the ratios of the shear pressures $\pi_j/\pi_i$. Even when $\pi_i$'s become
small compared with $e_i$'s, the ratios $\pi_j/\pi_i$ cannot vanish. $\eta_i$ is a coefficient
rather than shear viscosity of the species $i$ in the usual sense.
We now want to address the question whether and how dissipation in a multi-component fluid as a
whole can be described by a single effective shear viscosity coefficient.

Summing up the equations (\ref{dotpi}) over $i$, we obtain the hydrodynamic
equation for the total shear pressure $\pi=\sum_{i=1}^N \pi_i$:
\begin{equation}
{\dot \pi} = -\frac{2}{9}\frac{\pi e}{\eta_{eff}}
-\frac{4}{3}\frac{\pi}{\tau} +
\frac{8}{27}\frac{e}{\tau}
\label{dotpitotal}
\end{equation}
with the definition of the effective shear viscosity for a mixture via
\begin{equation}
\label{eta_eff1}
\frac{\pi e}{\eta_{eff}}=\sum_{i=1}^N \frac{\pi_i e_i}{\eta_i}
=\frac{7}{2} \sum_{i,j=1}^N \pi_i n_j \sigma_{ij} - 
\sum_{i,j=1}^N \pi_j n_i \sigma_{ij} \,.
\end{equation}
For the second identity the relation $T_i=e_i/(3n_i)$ and Eq. (\ref{eta_iso}) were used. $e$ is the
total local energy density. We recognize the equality of the two sums in (\ref{eta_eff1}) and thus
can write
\begin{equation}
\eta_{eff} = \frac{2}{5} e \left (
\sum_{i=1}^N \alpha_i \lambda_{mfp,i}^{-1} \right )^{-1} \,,
\label{eta_eff2}
\end{equation}
where $\lambda_{mfp,i}=1/\sum_{i=1}^N n_i \sigma_{ij}$ is the mean free path
of particles of species $i$ and $\alpha_i=\pi_i/\pi$ is the fraction of
the total shear pressure. For equal mean free paths, 
$\lambda_{mfp,i}=\lambda$, a multi-component system as a whole behaves
as if all particles were identical. For this special case one finds $\eta_{eff}=(2/5)(e/\lambda)$,
which is, indeed, exactly the result for a one-component system \cite{DeGroot,HM09}, as it should
be. For unequal mean free paths the effective shear viscosity additionally depends on the dynamic 
values $\alpha_i$. This means that in general, it is not possible
to describe a mixture by a free parameter $\eta_{eff}/s$. One has to
solve the equations (\ref{pi1}) for each shear pressure. $\eta/s$ is no longer a characteristic
value. Mean free paths and temperatures of all particle species are the relevant scales.

On the other hand, the shear viscosity $\eta$ of a mixture can be calculated
by means of the Green-Kubo formula \cite{Fuini:2010xz} and no dependence of $\eta$
on the ratios $\alpha_i$'s is expected. However, the Green-Kubo formula is applicable when
the particle system is in thermal equilibrium. Equation (\ref{eta_eff2}) gives a
more
general expression for the case the system is out of equilibrium. When the system is approaching
equilibrium, $\alpha_i$'s become constants, and their values can be obtained by solving
$\dot{\alpha}_i=0$ using Eqs. (\ref{dotpi}). The calculation is difficult, because we are dealing
with a system of non-linear equations. We can solve it for the simple case of $N=2$ and obtain:
\begin{equation}
\label{weight}
\left. \frac{\pi_1}{\pi_2} \right |_{\dot{\alpha_1}=\dot{\alpha_2}=0}
=\sqrt{\gamma^2+\frac{n_1}{n_2}}-\gamma \,,
\end{equation}
where
\begin{equation}
\gamma=\frac{5}{4n_2\sigma_{12}}\left ( \frac{1}{\lambda_{mfp,1}}-
\frac{1}{\lambda_{mfp,2}} \right ) +\frac{1}{2}\left ( 1-\frac{n_1}{n_2}
\right ) \,.
\end{equation}
Putting (\ref{weight}) into (\ref{eta_eff2}) gives the shear viscosity
of a two-component mixture in equilibrium, $\eta_{eq}$.
We would like to mention that the weights $\alpha_i$ in Eq. (\ref{eta_eff2}) depend
not only on the composition $n_1/n_2$, but also on the difference
between the inverses of the mean free paths, which is not trivial.

To verify our findings we now compare the solutions of hydrodynamic 
equations with those from kinetic transport calculations using the 
partonic cascade model BAMPS \cite{Xu:2004mz}. Previously published
works demonstrated that BAMPS results can be regarded as a benchmark 
for hydrodynamic calculations \cite{3rdO,Bouras:2010hm}.

As a further simplification we consider a spatially homogeneous 
two-component particle system. This means that terms containing the 
gradient $\partial_\mu u^\mu =1/\tau$ in Eqs. (\ref{dotpi}) and
(\ref{dotpitotal}) are dropped. Numerically this is realized by
confining particles in a static box.

Initial condition is chosen as follows: $T_1=T_2=400$ MeV, $n_1/n_2=5$,
$\pi_1/e_1=\pi_2/e_2=0.3$. Particles of species $1$ are
in chemical equilibrium, whereas particles of species $2$ are
undersaturated. Cross sections are $\sigma_{11}=3.88$ mb, 
$\sigma_{22}=\sigma_{11}/4$, and $\sigma_{12}=\sigma_{11}/2$.
We thus mimic elastic interactions among gluons (species $1$) and quarks (species $2$).

\begin{figure}[t!]
 \includegraphics[width=8.7cm]{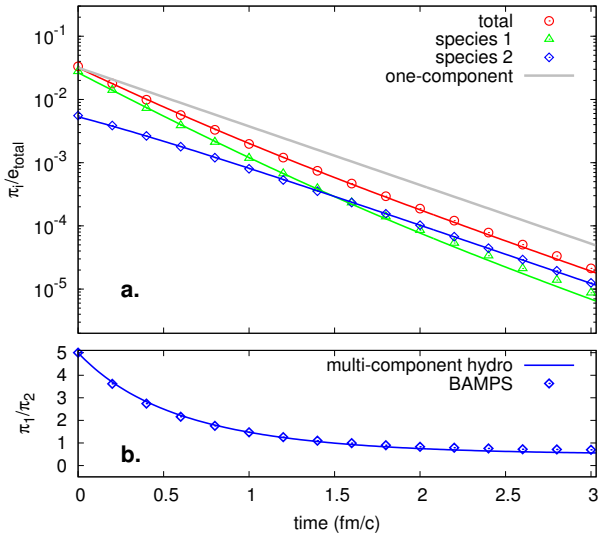}
\caption{ (Color online) (a) Time evolution of the partial and total shear pressures normalized by
total energy
density and (b) ratio of the partial shear pressures from BAMPS (symbols) and hydrodynamic
calculations (lines). Solid grey lines represent the effective one-component solution (see text).}
\label{fig:relax}
\end{figure}

Figure \ref{fig:relax}(a) shows the time evolutions of the partial shear pressures
$\pi_1$ and $\pi_2$, obtained by solving the hydrodynamic equations 
(\ref{dotpi}) (lines) and from BAMPS (symbols). Results are divided by total energy density. 
We see perfect agreement between the two approaches over up to $4$ orders of magnitude. This
demonstrates
the validity of the new multi-component hydrodynamic equations derived here.
To achieve such high numerical accuracy in BAMPS calculations we set the box
to have a size of $16\times 16 \times 16$ fm$^3$ and use a test 
particle number of $2\cdot 10^2$. The total particle number in one run
is $1.3\cdot 10^7$, and the results shown are the average obtained from $10^4$ runs.

Results for the total shear pressure divided by total energy density are also shown in 
Fig. \ref{fig:relax}(a), compared to solution of Eq. (\ref{dotpitotal})
with the shear viscosity in thermal equilibrium $\eta_{eq}$.
We see that the effective one-component hydrodynamic description 
underestimates the thermal equilibration at early times, while it
can be applied at late times when the system is approaching equilibrium.
In order to meet the BAMPS results at both initial and final times, one has to choose a smaller
$\eta_{eq}$. This
gives rise to conclusion that the extraction of the $\eta/s$ value
for the quark-gluon plasma at RHIC should be reexamined.

Finally, the $\pi_1/\pi_2$ ratio is depicted in Fig. \ref{fig:relax}(b).
Starting from the initial $\pi_1/\pi_2=5$, the ratio indeed relaxes to the value
calculated analytically, $\pi_1/\pi_2=0.488$. [see Eq. (\ref{weight})]. The effective relaxation
time for $\pi_1/\pi_2$ in this case is approximately $0.6$ fm/c and has a complicated dependence on
densities and cross sections.
\newline

In this Letter we have derived novel hydrodynamic 
equations for each component of a multi-component system. Although
the equations possess the relaxation form similar to that for 
a one-component system, they do reveal a new property -- the shear viscosity coefficients for each
component as well as the effective shear viscosity of the mixture depend on the ratios of shear
pressures. This leads to the conclusion that hydrodynamic behaviour of 
a multi-component system as a whole in general cannot be described by one-component hydrodynamic
equations with an effective shear viscosity coefficient calculated using standard methods (e.g.
Green-Kubo relations). It is inevitable to solve the hydrodynamic equations for 
each component. Instead of $\eta/s$, temperatures and mean free paths
are the relevant scales. We have confirmed our findings by comparing the solutions
of the derived hydrodynamic equations with those from kinetic transport 
calculations using BAMPS. Both results agree with each other with high accuracy.
Although the confirmation has been demonstrated for a static medium with
simplifications on cross sections, our findings are expected to hold
for general cases. A great potential of the present work lies in investigations of hydrodynamic
behaviour of the quark gluon plasma at RHIC and LHC \cite{El:prep}.
\newline

Authors are grateful to G. Denicol for
enlightening discussions. The BAMPS simulations were performed at the LOEWE-CSC Center. AE
acknowledges support by BMBF. AE and IB acknowledge support by HGS-HIRE and H-QM graduate
school.

This work has been supported by the Helmholtz International Center for FAIR within the
framework of
the LOEWE program launched by the State of Hesse.

\end{document}